\documentclass[prl,aps,twocolumn,raggedbottom]{revtex4}
\usepackage{latexsym}
\usepackage{amssymb}
\usepackage{amsmath}
\usepackage[varg]{txfonts}
\usepackage{mathrsfs}
\usepackage{upgreek}
\usepackage[small]{caption}
\usepackage[]{subfig}
\usepackage{verbatim}
\usepackage{array}
\usepackage{color}
\usepackage{graphicx}
\DeclareMathAlphabet{\mathpzc}{OT1}{pzc}{m}{it}
\usepackage{hyperref}
\usepackage{multirow}
\usepackage{natbib}

\begin{document}
\title{Study of the transition from pairing vibrational to pairing rotational regimes between magic numbers N=50 and N=82, with two--nucleon transfer}
\author{G. Potel}
\affiliation{Departamento de Fisica Atomica, Molecular y Nuclear, Universidad de Sevilla, Facultad de Fisica, Avda. Reina Mercedes s/n, Spain.}
\author{F. Marini}
\affiliation{Dipartimento di Fisica, Universit\`{a} di Milano,
Via Celoria 16, 20133 Milano, Italy.}
\author{A. Idini}
\affiliation{Dipartimento di Fisica, Universit\`{a} di Milano,
Via Celoria 16, 20133 Milano, Italy.}
\affiliation{INFN, Sezione di Milano Via Celoria 16, 20133 Milano, Italy.}
\author{F. Barranco}
\affiliation{Departamento de Fisica Aplicada III, Universidad de Sevilla, Escuela Superior de Ingenieros,
Sevilla, 41092 Camino de los Descubrimientos s/n,
Spain.}
\author{E. Vigezzi}
\affiliation{INFN, Sezione di Milano Via Celoria 16, 20133 Milano, Italy.}
\author{R. A. Broglia}
\affiliation{Dipartimento di Fisica, Universit\`{a} di Milano,
Via Celoria 16, 20133 Milano, Italy.}
\affiliation{INFN, Sezione di Milano Via Celoria 16, 20133 Milano, Italy.}
\affiliation{The Niels Bohr Institute, University of Copenhagen, Blegdamsvej 17,
2100 Copenhagen {\O}, Denmark.}
\begin{abstract}
Absolute values of two--particle transfer cross sections along the Sn--isotopic chain from closed shell to closed shell (${}^{100}$Sn,${ }^{132}$Sn) are calculated taking properly into account nuclear correlations, as well as the successive, simultaneous and non--orthogonality contributions to the differential cross sections.
The results are compared with systematic, homogeneous bombarding conditions ($p,t$) data. The observed agreement, almost within statistical errors and without free parameters, testify to the fact that theory is able to be quantitative in its predictions. Within this scenario, the predictions concerning the absolute value of the two--particle transfer cross sections associated with the excitation of the pairing vibrational spectrum expected around the the closed shell nucleus ${ }^{132}_{50}$Sn$^{ }_{82}$ can be considered quantitative. The same can be said to be true concerning the possibility of shedding light on the relative importance of successive and simultaneous transfer processes, through changes in the angular distributions expected to take place as a function of the bombarding energy (interference phenomena).

\end{abstract}

\maketitle

\section{Introduction}
Customary, the fingerprint of shell closure in nuclei is associated with a sharp, step-function--like distinction between occupied and empty single--particle states in correspondence with magic numbers (\cite{Mayer:55}, for a recent example see \cite{Cottle:10} and \cite{Jones:10}).

Away from closed shell, medium-heavy nuclei become, as a rule, superfluid, the distinction between occupied and empty states being blurred within a 2--3 MeV energy interval centered around the Fermi energy. This phenomenon is clearly captured by the Bogoliubov--Valatin quasiparticle transformation \cite{Bogoliubov:58,Valatin:58}.

\begin{equation}
 \alpha^{\dagger}_{\nu} = U_{\nu}a^{\dagger}_{\nu} - V_{\nu}a_{\bar{\nu}} \;.
\end{equation}
 It provides the rotation in Hilbert space of creation and annihilation fermion operators $a^{\dagger},a$  which diagonalizes the mean field pairing Hamiltonian in the state \cite{Bardeen:57a,Bardeen:57b},
\begin{equation}
 \vert BCS \rangle \sim \prod_{\nu>0} \alpha_{\nu} \alpha_{\bar{\nu}} \vert 0 \rangle \sim \prod_{\nu>0} (U_{\nu}+V_{\nu}a^{\dagger}_{\nu}a^{\dagger}_{\bar{\nu}})\vert 0 \rangle.
\label{eq.BCS}
\end{equation}
This state is a wavepacket in the number of pairs. A consequence of this fact is that the pair creation and annihilation operators
\begin{equation}
 P^{\dagger} = \sum_{\nu>0} a^{\dagger}_{\nu}a^{\dagger}_{\bar{\nu}}, \qquad P^{} = \sum_{\nu>0} a^{}_{\bar{\nu}}a^{}_{\nu},
\end{equation}
display a finite average value in it (condensed Cooper pair field),
\begin{equation}
 \alpha_0=e^{i\phi}\alpha_0'=\langle BCS \vert P^{\dagger} \vert BCS \rangle =  \langle BCS \vert P^{} \vert BCS \rangle,
\end{equation}
where $\alpha'_0=\sum_{\nu>0}U_\nu V_\nu$ (the pairing gap $\Delta$ being $G\alpha_0$). The $|BCS\rangle$ state thus defines
 a privileged orientation in gauge space. $\phi$ being the gauge angle between the laboratory and the intrinsic, body--fixed--frame of reference. The associated emergent property corresponding to generalized rigidity with respect to pair transfer.

\begin{figure}
	\begin{center}
		\includegraphics[width=0.37\textwidth]{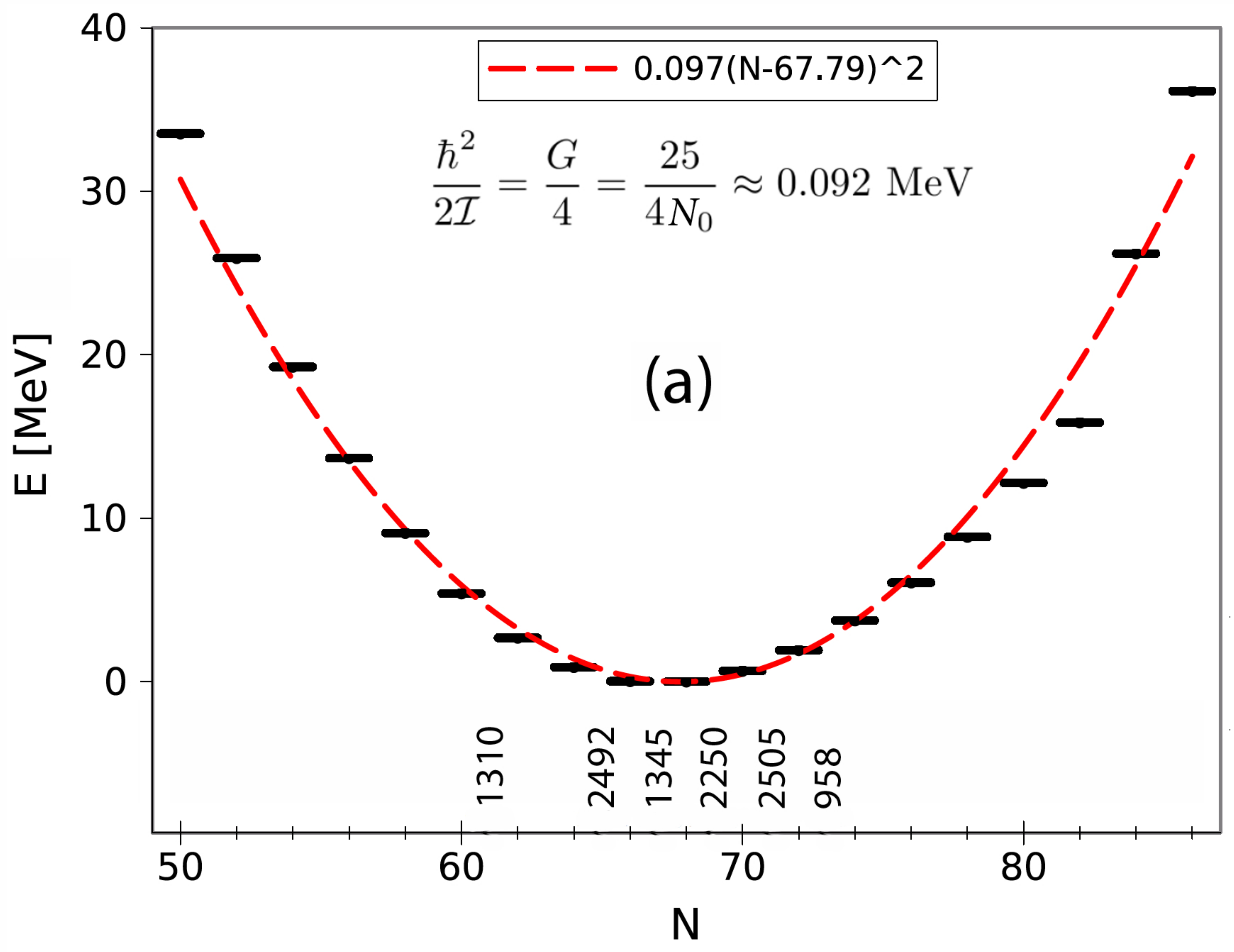}\\
	\end{center}
	\begin{center}
		\includegraphics[width=0.36\textwidth]{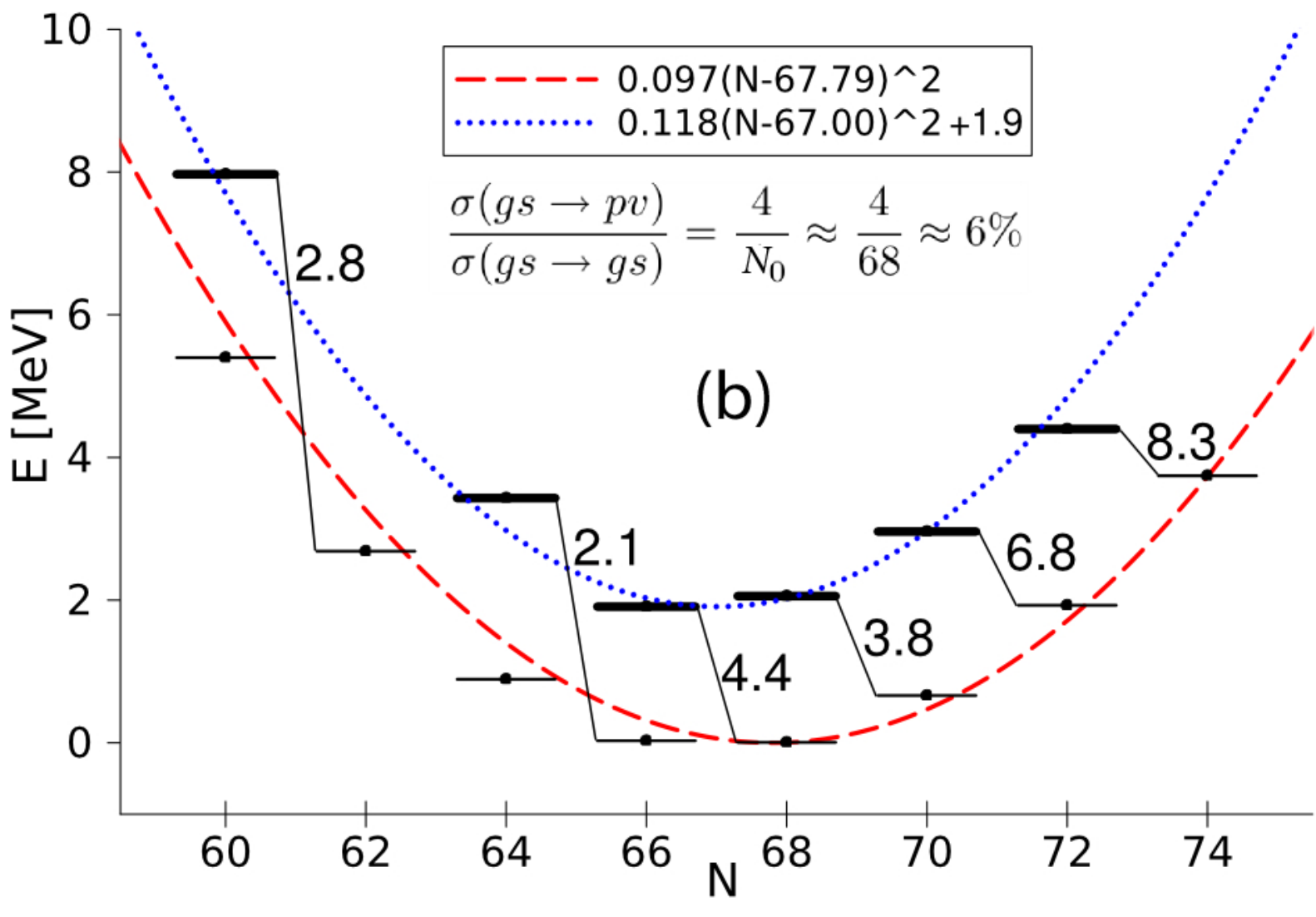}
	\end{center}
	\caption{Experimental energies of the $J^\pi = 0^+$ states of the Sn--isotopes ((a) ground state, (b) excited states), populated in $(p,t)$ reactions [11--14]. The heavy drawn horizontal lines represent the values of the expression $E=-B({}^A_{50} \textrm{Sn}_N) + E_\textrm{exc} + 8.124 A \;\textrm{MeV} $ $ + 46.33 \;\textrm{MeV}$, where $B({}^A_{50} \textrm{Sn}_N)$ is the binding energy of the Sn--isotopes of mass number $A$, and $E_\textrm{exc}$ is the weighted (with $\sigma(0_i^+)$) average energy of the excited $0^+$ states below 3 MeV. The dashed and dotted lines represent the parabola's given in the insets of (a) and (b), corresponding to the ground state and to the (average) excited state--based pairing rotational bands. The numbers labelling in (b) the thin lines connecting members of the ground state and of the excited state pairing rotational bands are the relative (in \%) $(p,t)$ cross sections $\sum_i \sigma(gs\rightarrow 0^+_i)$ normalized with respect to the ground state cross sections. The absolute values of these cross sections (in $\mu$b units) [11--14] are given in (a) along the abscissa. Simple estimates of the moment of inertia and of the cross talk expected between excited $(pv)$ and $(gs)$ pairing rotational bands obtained making use of the single $j$--shell model are also shown (see e.g. [10], app. H).}\label{fig1}
\end{figure}

 Taking into account the correlations among quasiparticles induced by fluctuations in $\alpha_0$ (gauge angle) induced by the field $(P^+ - P)$, leads to symmetry restoration. That is, to pairing rotations (Fig.\ref{fig1} (a), \cite{Anderson:58,Bes:66} see also \cite{Brink:05}, Sect. 6.6 and App. I).

 Taking into account the fluctuation in $\alpha_0^\prime$ induced by the field $(P^+ + P)$ leads to two--quasiparticle--like states called pairing vibrations,lying on top of twice the pairing gap \cite{Anderson:58,Bes:66}. In the superfluid case these vibrations are little collective (see Fig. \ref{fig1}(b), pairing rotational bands based on pairing vibrational excitations displaying a few \% cross section as compared to the $gs \rightarrow gs$ transitions).

 These fluctuations are, of course, already present in the normal ground state of closed shell nuclei in which case, due to the possibility of distinguishing between occupied and empty states, they are quite collective \cite{Bes:66}. In other words, in closed shell nuclei ($\alpha_0=0$), the quantity
\begin{multline}
 \sigma=\langle (\alpha-\alpha_0)^2 \rangle^{1/2} = \left[ \left( \langle 0 \vert P^{\dagger} P\vert 0 \rangle + \langle 0 \vert P^{} P^{\dagger} \vert 0 \rangle\right) /2 \right]^{1/2}\\
 = \left[ \sum_{i} \left( \vert \langle i(A_0-2)\vert P \vert 0 \rangle \vert^{2} + \vert \langle i(A_0+2)\vert P^{\dagger} \vert 0 \rangle \vert^{2} \right) /2 \right]^{1/2},
\end{multline}
where $|0\rangle = | gs(A_0) \rangle$, $\sigma$ displays finite values, of the order of $E_{corr}/G\approx 1$ MeV$/G\approx 5-10$, $E_{corr}$ being the average correlation energy of e.g. two neutrons (two neutron holes) outside (in) the closed shell system of mass number $A_0$. Of notice that the marked deviations observed around $N=50$ and $N=82$ in the energies of the Sn--ground states as compared with the parabolic fitting (see Fig. \ref{fig1}(a)), is associated with the fact that in these cases we have to deal with pair vibrations of a normal nucleus and not with pairing rotations of a superfluid system.
As emerges from (3), (4) and the paragraph following this equation, as well as (5), one can posit that two--nucleon transfer is the specific probe of pairing in nuclei.

\begin{figure}
	\begin{center}
		\includegraphics[width=0.37\textwidth]{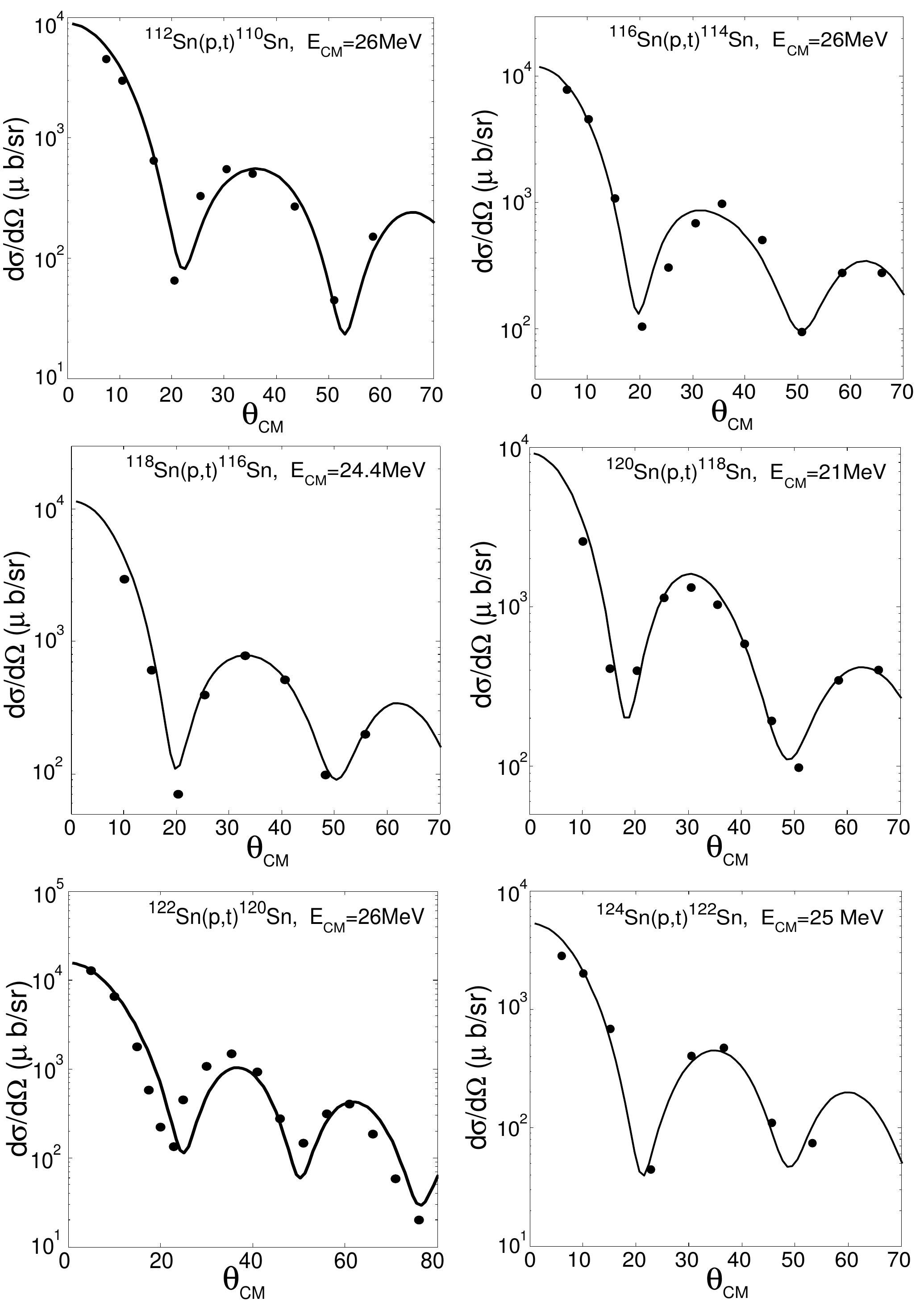}
	\end{center}
	\caption{Absolute calculated cross section predictions in comparison with the experimental results of \cite{Guazzoni:99,Guazzoni:04,Guazzoni:06,Guazzoni:11}}\label{fig2}
\end{figure}

From the above narrative, and from the definition of the (nuclear) correlation length $\xi=\hbar v_F/2 \delta$, where $\delta=G\alpha_0'$ in the case of superfluid nuclei, and $\delta=E_{corr}$, in the case of closed shell nuclei, Cooper pair partners are paired over distances considerably larger than nuclear dimensions ($\xi \approx 20-30$ fm). This estimate, together with the fact that $(gs)\rightarrow(gs)$ two--particle transfer cross section can be written as,
\begin{equation}
 \sigma (gs \rightarrow gs) \sim\left\{
\begin{array}{l l}
  \alpha_{0}^{2} & \quad (\alpha_0 \neq 0),\\ \\
  \sigma^{2} & \quad (\alpha_0 = 0),\\ \end{array} \right.
\end{equation}
implies that Cooper pair partners can remain correlated across regions of the system for which $G(x)=0$ (a result first realized in connection with the Josephson effect \cite{Josephson:62}).

Consequently, and exception made for the closing of single--particle channels due to $Q$--value effects (see below), one expects that successive transfer induced by the single--particle mean field,
\begin{equation} \label{eq:ur}
	U(r) = \int \textrm{d}^3 r' \rho (r') V_{np} \left( |\vec{r} - \vec{r'}| \right) \;,
\end{equation}
will be dominant over simultaneous transfer, let alone over transfer induced by the pairing interaction. The proton--neutron interaction $V_{np}$ appearing in (\ref{eq:ur}) has been parametrized according to \cite{Tang:65}, its strength adjusted so as to reproduced the intermediate channel deuteron binding energy,

Making use of the two-particle transfer spectroscopic amplitudes $B_\nu=(j_\nu+1/2)^{1/2}U_\nu V_\nu$ \cite{Broglia:73} of the Cooper pair condensate under study, and standard optical parameters \cite{Guazzoni:99,Guazzoni:04,Guazzoni:06,Guazzoni:11}, the absolute differential cross sections associated with the reactions  $^{A}$Sn$(p,t)^{A-2}$Sn(gs) ($A$=124,122,120,118,116 and 112) were calculated. In all cases, successive, simultaneous and non--orthogonality contributions (post representation) to the cross section were considered (see \cite{Gotz:75,Igarashi:91} and refs. therein, see also \cite{Broglia:00b}). The results, in comparison with the experimental data \cite{Guazzoni:99,Guazzoni:04,Guazzoni:06,Guazzoni:11}, are displayed in Fig. \ref{fig2}. Theory provides, without any free parameters, an account of the absolute value of all measured differential cross sections within limits well below the (estimated) 15\% (systematic) experimental error, and almost within statistical errors.

\begin{figure}
	\begin{center}
		\includegraphics[width=0.40\textwidth]{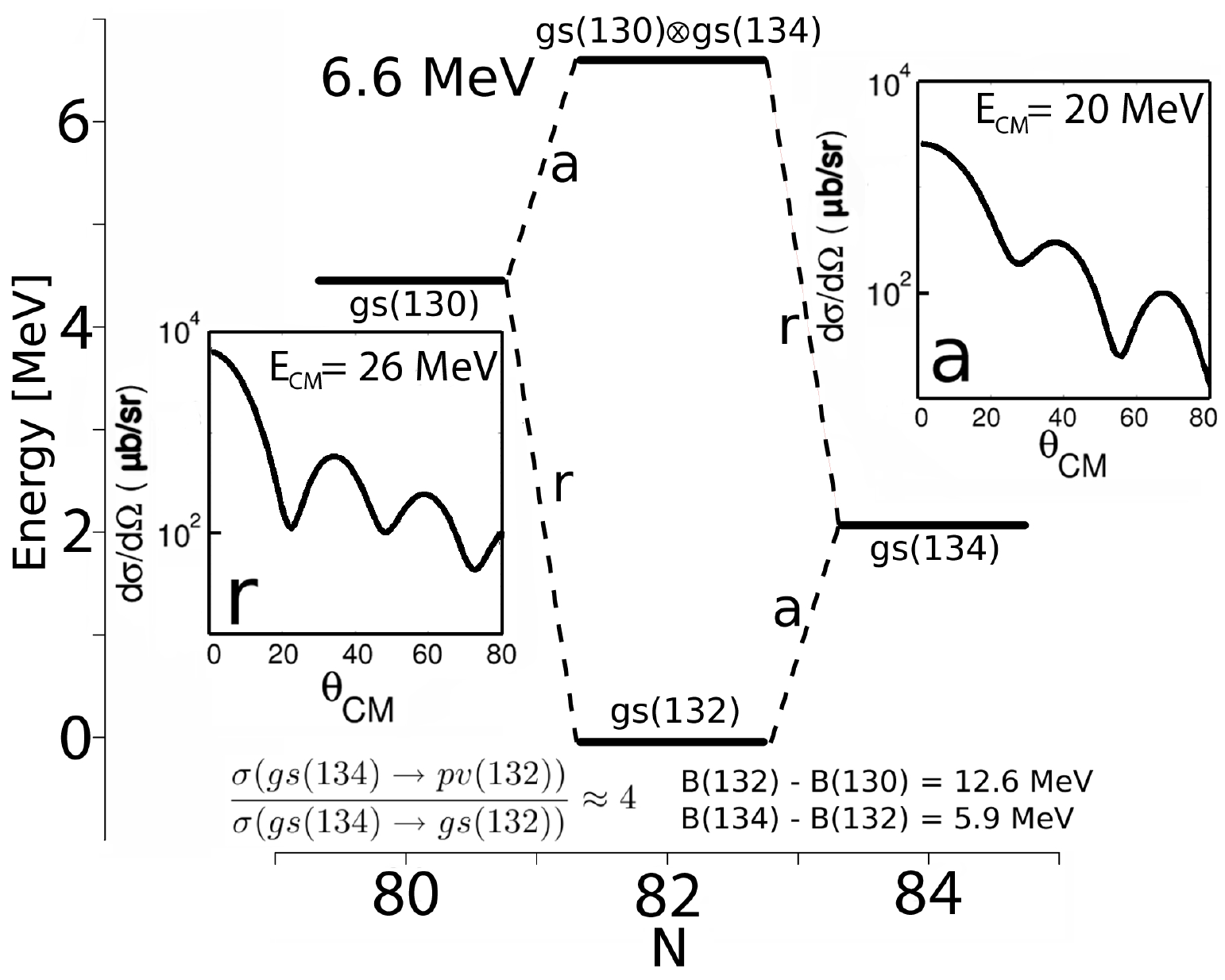}
	\end{center}
	\caption{Pairing vibrational spectrum around ${}^{132}$Sn. The absolute differential cross sections associated with the reactions ${}^{132}$Sn$(t,p){}^{134}$Sn$(gs)$ (pair addition: a) and ${}^{132}$Sn$(p,t){}^{130}$Sn$(gs)$ (pair removal: r) at $E_{CM}=20$ MeV and 26 MeV respectively are also displayed.}\label{fig3}
\end{figure}
In keeping with these results we present below predictions concerning the pairing vibrational spectrum of the closed shell nucleus $^{132}_{50}$Sn$_{82}$ and the associated absolute differential cross sections. Within the harmonic approximation \cite{Bes:66,Broglia:73}, the two phonon pairing vibrational $0^+$ state is predicted at an excitation energy of 6.6 MeV (see Fig. 3). At variance with the superfluid (pairing rotational) case (see Fig. \ref{fig1}(b)), this excited $0^+$ state is expected to be populated with a large cross section as compared to the gs$\rightarrow$gs transition, a direct consequence of the clear distinction which can be operated between occupied ($V^2 \approx 1, U^2 \approx 0$) and empty ($V^2 \approx 0, U^2 \approx 1$) states around closed shell systems.

\begin{figure}[h!]
	\begin{center}
		\includegraphics*[width=0.37\textwidth]{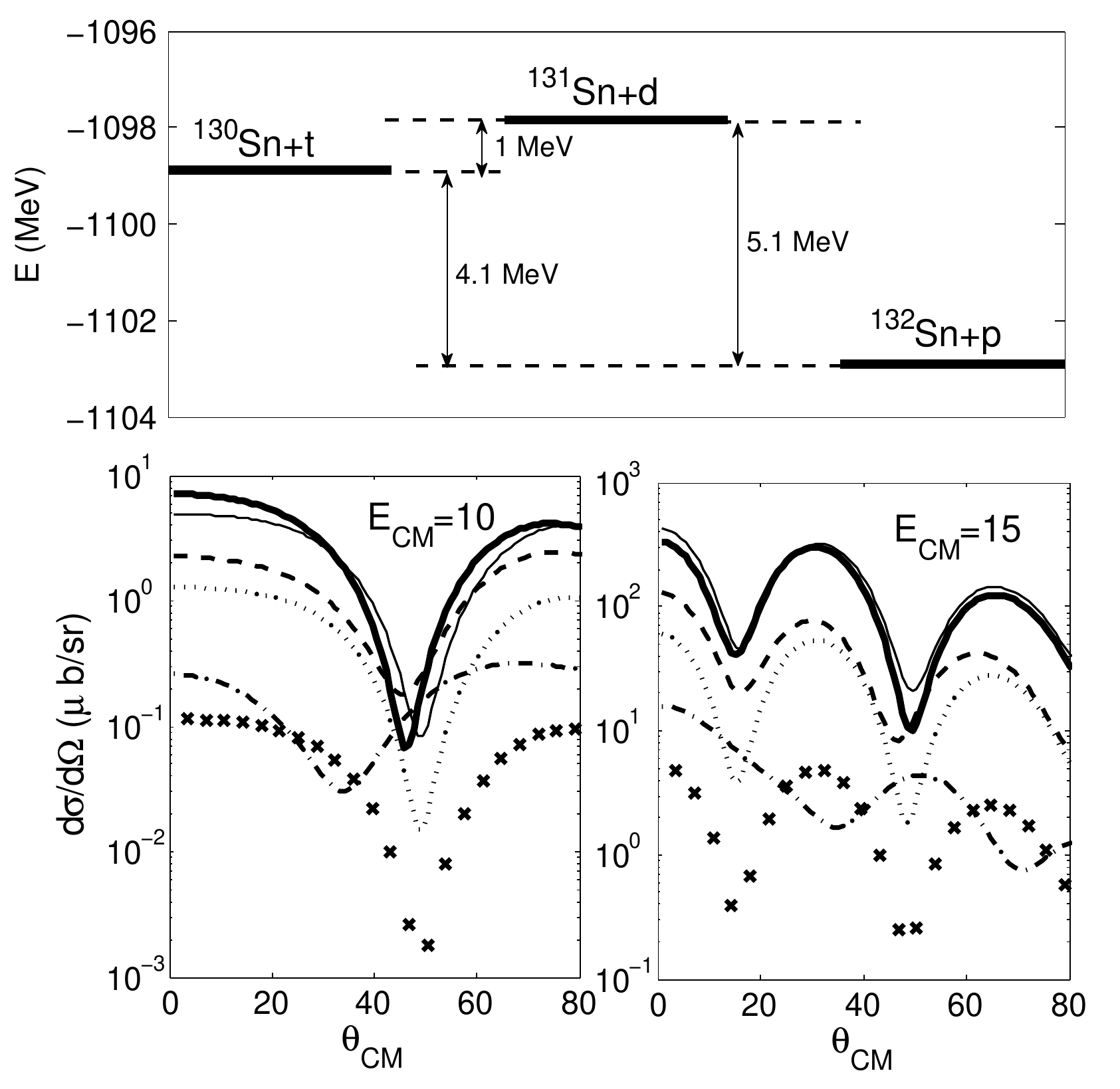}
	\end{center}
	\caption{Absolute differential cross section associated with the reaction ${}^{132}$Sn$(p,t){}^{130}$Sn$(gs)$ at two CM bombarding energies (in MeV). Successive (thin continuous curve), simultaneous (dotted), non--orthogonality (dashed), simult.+non orth. (dotted--dashed), pairing force (crosses) and total (thick continuous curve) cross sections are also displayed. In the upper part of the figure the level scheme associated with the initial, final and lowest energy intermediate channels are shown.}\label{fig4}
\end{figure}

While the bombarding conditions used in these calculations are similar to those encountered in connection with the experimental data shown in Fig. \ref{fig2} \cite{Guazzoni:99,Guazzoni:04,Guazzoni:06,Guazzoni:11}, the inverse kinematics techniques required when dealing with $^{132}$Sn will pose severe limitations to such (optimal) choices. It turns out that the respect of such limitations may contain the key for a qualitative advance in the understanding of the two--nucleon transfer reaction mechanism at large, similar, with all required caveats, to that which took place in the understanding of pair tunneling phenomena in connection with the Josephson effect (see e.g. \cite{Cohen:62} and refs. therein).

This is in keeping with the fact that there exists an important ($Q$--value, kinematic--like) difference between the pairing coupling scheme expected around and away from closed shells. In fact, while the binding energy of open shell superfluid isotopes is a smoothly varying function of mass number, the situation is quite different around closed shell ($A_0$). Both the $A_0$ system, as well as the ($A_0 \pm 2$) nuclei are well bound, the associated $Q$--value being, as a rule, rather unfavorable for single--particle transfer, let alone for twice such a process (successive).

\begin{figure}
	\begin{center}
		\includegraphics*[width=0.32\textwidth]{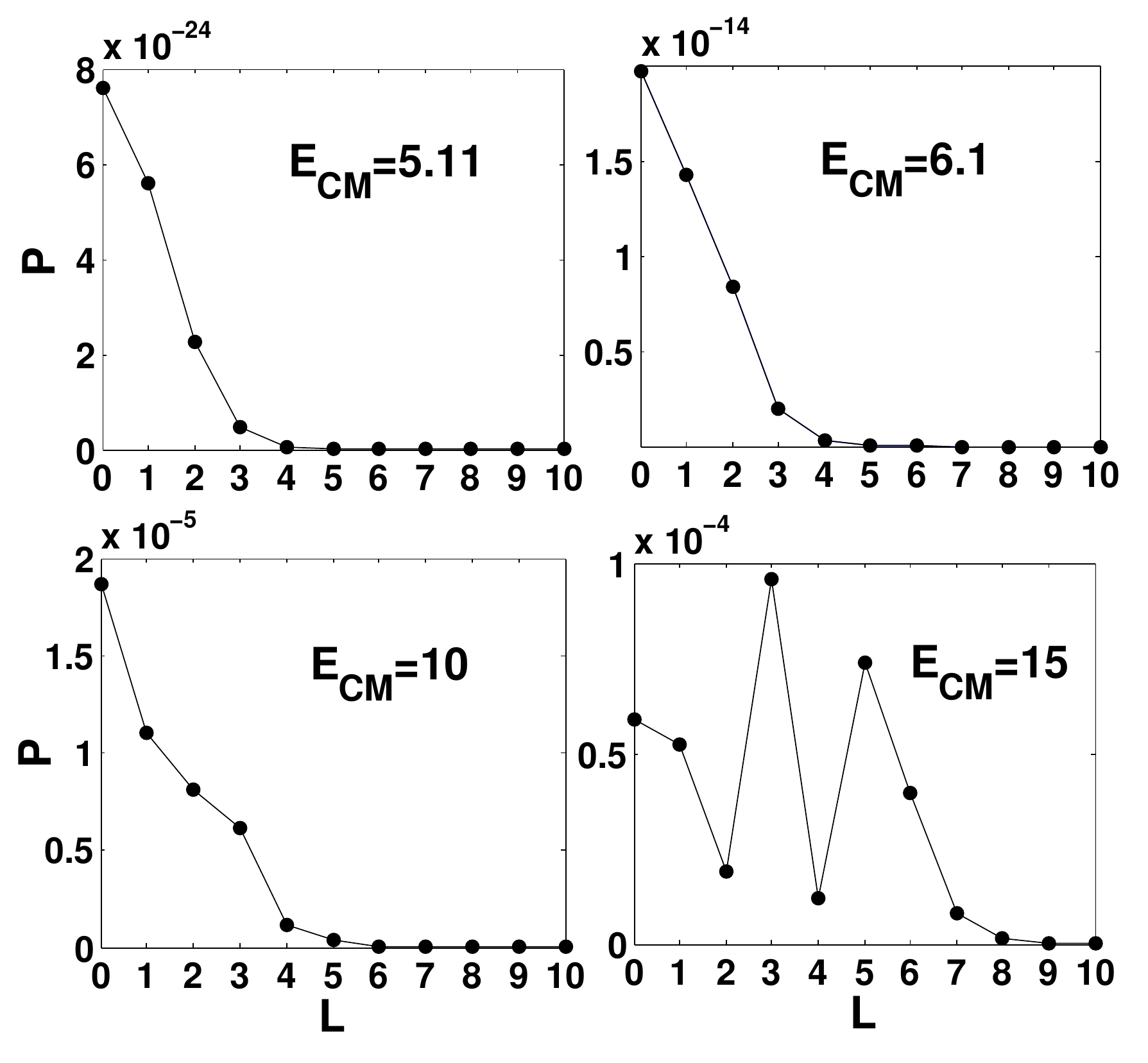}
	\end{center}
	\caption{Probability with which each partial wave $L$ contributes to the (total) two--particle transfer cross section ${}^{132}$Sn$(p,t){}^{130}$Sn$(gs)$ (see Table \ref{table1}).}\label{fig5}
\end{figure}

\begin{table}
\begin{center}
\begin{small}
\begin{tabular}{|l|c c c c|}
\hline
& \multicolumn{4}{c|} {$\sigma (\mu \text{b})$}\\
\cline{2-5}
& 5.11 MeV&  6.1 MeV &  10.07 MeV& 15.04 MeV  \\
\hline
total & $1.29 \times10^{-17}  $  &$3.77 \times10^{-8}$& $39.02 $& 750.2  \\
\cline{1-1}
successive & $9.48 \times10^{-20}$ & $1.14 \times10^{-8}$ & $44.44 $ & 863.8   \\
\cline{1-1}
simultaneous &$1.18 \times10^{-18}$& $8.07 \times10^{-9}$ & 10.9 & 156.7  \\
\cline{1-1}
non--orthogonal &$2.17 \times10^{-17}$ & $7.17 \times10^{-8}$ & 22.68   & 233.5 \\
\cline{1-1}
non--orth.+sim. &$1.31 \times10^{-17}$ & $3.34 \times10^{-8}$ &  3.18  & 17.4 \\
\cline{1-1}
pairing &$1.01 \times10^{-19}$ & $6.86 \times10^{-10}$ &  0.97  & 14.04 \\
\hline
\end{tabular}
\end{small}
\caption{Absolute value of the cross sections displayed in Fig. \ref{fig4} (plus those associated with $E_{CM}=5.11$ and 6.1 MeV), integrated over the range $0^\circ \leq \theta_{\textrm{CM}} \leq 80^\circ$.}\label{table1}
\end{center}
\end{table}

This is clearly seen from the (bombarding) energy dependence of the absolute cross sections associated with the reaction $^{132}$Sn$(p,t){}^{130}$Sn (Fig. \ref{fig4} and Table \ref{table1}). Also reported in Fig. \ref{fig4} is a simple estimate of the contribution to the differential cross section arizing from the two--body pairing interaction, in which case the associated transfer field can be written as $V_P = \sigma (P^\dagger +P)$. As observed from the level scheme displayed on top of this figure, below 4.1 MeV no (real) two--particle transfer can take place. One needs to reach $CM$ bombarding energies of the order of 10 MeV, to obtain values of the absolute cross section which are barely observable, as a result of the cancellation taking place between simultaneous and non--orthogonality contributions and of the (hindered) $Q$--value dependence of successive transfer. By properly tuning the bombarding conditions, one can reduce the role successive transfer plays in the process, and thus change the shape and absolute value of the two--particle transfer differential cross section. This is a direct consequence of the very low number of partial distorted waves controlling the transfer process at these threshold energies (see Fig. \ref{fig5}).

From the analysis of systematic two--particle transfer data on open shell Sn--isotopes, with the help of a unified nuclear structure--reaction mechanism description \cite{Brink:05,Broglia:00b}, it is concluded that theory is able to account for the experimental findings within experimental errors and without free parameters. Applying the same theoretical tools to the pairing vibrational coupling scheme expected around the closed shell nucleus $^{132}_{50}$Sn$_{82}$, likely opens the possibility of shedding light, by accurately tuning the bombarding conditions, on the mechanism which is at the basis of two--nucleon transfer reactions, as well as to set a lower limit to the contribution of the, likely highly dressed, nucleon--nucleon pairing interaction.

We also acknowledge discussions with Luisa Zetta and Paolo Guazzoni on one of the most systematic two--nucleon transfer studies of recent date available in the literature, which is at the basis of the present letter.

\bibliographystyle{apsrev}

\end{document}